\documentstyle[prl,multicol,aps]{revtex}
\begin{document}

\input{epsf.tex}
\epsfverbosetrue

\title{Nonlinear Modes of a Macroscopic Quantum Oscillator}

\author{Yuri S. Kivshar$^1$, Tristram J. Alexander$^1$,
and Sergey K. Turitsyn$^2$}

\address{$^1$ Optical Sciences Center, Research School of Physical
Sciences and Engineering, Institute of Advanced Studies\\
 Australian National University, Canberra ACT 0200, Australia \\
$^2$ Division of Electrical Engineering and Computer Science, Aston
University, Birmingham B4 7EK, UK}

\maketitle

\begin{abstract}
We consider the Bose-Einstein condensate in a parabolic trap as
{\em a macroscopic quantum oscillator} and describe, analytically and
numerically, its collective modes--a nonlinear generalisation
of the (symmetric and antisymmetric) Hermite-Gauss eigenmodes
of a harmonic quantum oscillator.
\end{abstract}


\begin{multicols}{2}
\narrowtext

The recent observation of different types of Bose-Einstein condensation
(BEC) in atomic clouds \cite{exper} led to the foundation of a new direction
in the study of macroscopic quantum phenomena.  From a general point of
view, the
dynamics of gases of cooled atoms  confined in a magnetic trap  at very
low temperatures can be described by an effective equation for the
condensate wave function known as the Gross-Pitaevskii (GP) equation
\cite{review}. This is a classical  nonlinear equation that takes into
account the effects of the particle interaction through an effective mean
field, and therefore it can be treated as {\em a nonlinear generalization} of a
text-book problem of quantum mechanics, i.e. as {\em a macroscopic quantum
oscillator}.

Similar models of the confined dynamics of macroscopic quantum systems
appear in other fields, e.g. in the case of an electron gas confined in  a
quantum well \cite{el}, or optical modes in a photonic microcavity
\cite{ph_atom}. In all such systems, confined single-particle states are
restricted to a set of discrete energies that form a set of eigenmodes. A
classical and probably most familiar example of such a system is a harmonic
quantum oscillator with equally spaced energy levels \cite{landau}.

When, instead of single-particle states, we describe quasiclassically
a system of interacting bosons in a macroscopic ground state confined in an
external potential,
a standard application of the mean-field theory allows us to introduce a
macroscopic wave function as a classical field  $\Psi ({\bf R},t)$ having
the meaning of the order parameter. The equation for the function $\Psi ({\bf
R},t)$ looks
similar to that of a single-particle oscillator, but it also includes the
effect of interparticle interaction, taken into account as a mean-field
nonlinear term. Then, the important questions are: {\em Does the physical
picture of eigenmodes remain valid in the nonlinear case,
and what is the effect of nonlinearity on the modes ?}  In
this paper we analyse {\em nonlinear eigenmodes of a macroscopic quantum
oscillator} as a set of nonlinear stationary states that extend the
well-known Hermite-Gauss eigenfunctions. We also make a link between
seemingly different approximations,  the well-known Thomas-Fermi
approximation and the perturbation theory developed here for the case of 
weak nonlinearity. For both attractive and repulsive 
interaction, we demonstrate a close connection between the nonlinear modes
and (bright and dark) multi-soliton stationary states.

We consider the macroscopic dynamics of condensed atomic clouds
in a three-dimensional, strongly anisotropic, external parabolic potential
created by a magnetic trap. The BEC collective dynamics can be described
by the GP equation,
\begin{equation}
\label{GPeq}
i\hbar \frac{\partial \Psi}{\partial t} = - \frac{\hbar^{2}}{2m}
\nabla^{2} \Psi + V({\bf R})\Psi + U_{0}|\Psi|^{2}\Psi,
\end{equation}
where $\Psi({\bf R},t)$ is the macroscopic wave function of
a condensate, $V({\bf R})$ is a parabolic trapping potential,
and the parameter $U_{0} = 4\pi \hbar^{2}(a/m)$ characterises the
two-particle interaction proportional to the s-wave scattering length
$a$.  When $a>0$, the interaction between the particles in the condensate
is  {\em repulsive}, whereas for $a<0$ the interaction is {\em attractive}.
In fact, the scattering length $a$ can be continuously detuned from
positive to negative values by varying the external magnetic field near the
so-called Feshbach resonances \cite{feshbach}.

First of all, we derive from Eq. (\ref{GPeq}) an effective one-dimensional
model, assuming the case of a highly anisotropic (cigar-shaped) trap of the
axial symmetry  $ V({\bf R}) = \frac{1}{2} m\omega_{\perp}^2 (R_{\perp}^2
+\lambda X^2)$, where $R_{\perp} = \sqrt{Y^2 + Z^2}$.  This means that
$\lambda \equiv \omega_{\parallel}^{2}/ \omega_{\perp}^{2} \ll 1$, and the
transverse structure of the condensate, being close to a Gaussian in shape,
is  mostly defined by the trapping  potential \cite{victor}.

Measuring the spatial variables in the units of the longitudinal  harmonic
oscillator length $a_{ho}=(\hbar/m\omega\sqrt{\lambda})^{1/2}$, and  the
wavefunction amplitude, in units of $(\hbar\omega_{\perp}
/2U_{0}\sqrt{\lambda})^{1/2}$, we  obtain the following dimensionless
equation:
\begin{equation}
\label{Phieq}
i\frac{\partial \Psi}{\partial t} + \nabla^2 \Psi
- [\lambda^{-1}(y^2 + z^{2}) +x^2]\Psi + \sigma|\Psi|^{2}\Psi = 0,
\end{equation}
where time is measured in the units of $(2/\omega_{\perp} \sqrt{\lambda})$,
$(x, y, z) =(X, Y, Z)/a_{ho}$, and the sign $\sigma = {\rm sgn}(a) = \pm 1$
in front of the nonlinear  term is defined by the sign of the s-wave
scattering length of the two-body  interaction.

We assume that in Eq. (\ref{GPeq}) the nonlinear interaction is weak
relative to the trapping potential force in the transverse dimensions, i.e. 
$\lambda \ll 1$.  Then, it follows from Eq. (\ref{Phieq})  that  the transverse
structure of the condensate is of order of $\lambda$, and the  condensate
has a cigar-like shape.  Therefore, we can look for solutions of  Eq.
(\ref{Phieq}) in the form,
\[
 \Psi(r,x,t) = \Phi(r) \psi(x,t) e^{-2i\gamma t},
\]
where $r= \sqrt{y^2 + z^2}$, and $\Phi(r)$ is a solution of the auxiliary
problem for the 2D radially symmetric quantum harmonic oscillator
\[
\nabla_{\perp}^2 \Phi + 2 \gamma \Phi - (r^{2}/\lambda)\Phi = 0,
\]
which we take in the form of the no-node ground state,
$ \Phi_{0}(r) = C \exp (- \gamma r^2/2)$, where $\gamma =
1/\sqrt{\lambda}$.  To preserve all the information about the structure of
the 3D condensate in an asymmetric trap describing its properties by the
effective GP equation for the longitudinal profile, we impose the
normalisation for $\Phi_0(r)$ that yields $C^{2} =  \gamma/\pi$.

After substituting such a factorized solution into Eq. (\ref{Phieq}),
dividing by $\Phi$ and integrating over the transverse cross-section of the
cigar-shaped condensate, we finally derive the  following 1D {\em
nonstationary} GP equation
\begin{equation}
\label{GPeq_1}
i \frac{\partial \psi}{\partial t} + \frac{\partial^{2}\psi}{\partial x^{2}} 
- x^2 \psi + \sigma |\psi|^{2} \psi = 0.
\end{equation}

The number of the condensate particles $N$ is now defined as
$N = (\hbar\omega/2U_{0}\sqrt{\lambda})Q$, where
\begin{equation}
\label{Q_eq}
Q = \int^{\infty}_{-\infty} |\psi(x,t)|^{2}d x
\end{equation}
is the integral of motion for the normalised nonstationary GP equation
(\ref{GPeq_1}).

Equation (\ref{GPeq_1}) includes all the terms of the same order, and it
describes {\em a longitudinal profile} of the condensate state in a highly
anisotropic trap.  In the linear limit, i.e. when formally
$\sigma \rightarrow 0$, Eq. (\ref{GPeq_1}) becomes  the well-known
equation for a harmonic quantum oscillator.  Its stationary localised
solutions,
\begin{equation}
\label{local}
\psi(x,t) = \phi(x) e^{- i\Omega t},
\end{equation}
 exist  only for discrete values of $\Omega$,  such that
\[
\Omega_{n} = 1+2n, \;\;\; n=0,1,2,\ldots,
\]
and they are defined
through the Hermite-Gauss polynomials, $\phi_{n}(x) = c_{n}{\rm
e}^{-x^{2}/2}H_{n}(x)$, where $c_n = (2^n n! \, \sqrt{\pi})^{-1/2}$, and
\begin{equation}
\label{Hermite}
H_{n}(x) = (-1)^{n}{\rm e}^{x^{2}/2}\frac{d^{n}({\rm e}^{-x^{2}/2})}{dx^{n}},
\end{equation}
so that $H_{0} = 1$, $H_{1} = 2x$, etc.

In general, the localised solutions of Eq. (\ref{GPeq_1}) for $\sigma \neq 0$
can be found
only  numerically.  All such solutions can be characterised by the
dependence of the invariant (\ref{Q_eq}) on the effective nonlinear frequency
$\Omega$. However, in some particular limits  we can employ different
approximate methods to find the localised solutions analytically.

 First of all, to describe the effect of {\em weak nonlinearity}, we use
the perturbation theory based on the expansion of the general solution of
Eq. (\ref{GPeq_1}) in the infinite set of the eigenfunctions
(\ref{Hermite}).  A similar approach has been used earlier in the theory
of the dispersion-managed optical solitons \cite{turitsyn}. To apply such a
perturbation theory, we look for solutions of Eq. (\ref{GPeq_1}) in the form
[cf. Eq. (\ref{local})]
\begin{equation}
\label{expansion}
\psi (x, t) = e^{-i\Omega t} \sum_{n=0}^{\infty} B_n \phi_n(x),
\end{equation}
where $\phi_n(x)$ are the eigenfunctions of the linear equation for a
harmonic oscillator that satisfy the equation
\begin{equation}
\label{exp_phi}
\frac{d^2\phi_n}{dx^2} - x^2 \phi_n + \Omega_n \phi_n =0.
\end{equation}
 Inserting the expansion (\ref{expansion}) into Eq. (\ref{GPeq_1}),
multiplying by $\phi_n$ and averaging, we obtain a system of algebraic
equations for the coefficients
\begin{equation}
\label{station}
(\Omega - \Omega_m) B_m - \sigma \sum_{n,l,k} V_{m,n,l,k} B_n B_l B_k = 0,
\end{equation}
where
\[
V_{m,n,l,k} = \int_{-\infty}^{+\infty} \phi_m(x) \phi_n(x) \phi_l(x)
\phi_k(x) dx.
\]
Equation (\ref{station}) can be also rewritten in the traditional form
$\delta (H + \Omega Q) =0$, and it allows us to develop a perturbation theory
for small nonlinearities, for any Hermite-Gaussian eigenmode. For example, let
us consider the ground state mode at $n=0$. Assuming $B_0 \gg B_m$ for $m
\neq 0$ and the condition of the symmetric solution $\phi(x) = \phi(-x)$,
i.e. $B_{2k+1} =0$ for any $k$, we find the corrections, $\Omega \approx
\Omega_0 + \sigma V_{0,0,0,0} |B_0|^2$, and
\[
B_{2k} = \frac{\sigma V_{2k,0,0,0}}{(\Omega - \Omega_{2k})} |B_0|^2 B_0,
\;\; k \neq 0.
\]
This allows us to calculate the asymptotic expansion of the invariant $Q$ for
small nonlinearities,
\begin{equation}
\label{as}
Q = \sum_k|B_{2k}|^2  \approx - \sigma a_0 (\Omega -\Omega_0) [ 1 + b_0
(\Omega -\Omega_0)^2],
\end{equation}
where the coefficients are:
 \[
a_0 = \sqrt{2\pi}, \;\;\;  b_0 = \sum_{k=1}^{\infty} \frac{(2k) !}{(4k)^2
(k! \, 2^{2k})^2}.
\]
Higher-order modes can be considered in a similar way, and the results 
are similar to Eq. (\ref{as}) where $a_{0}$ and $b_{0}$ change to $a_{n}$ 
and $b_{n}$ respectively, where $a_{n}$ and $b_{n}$ depend on the mode order.

In the opposite limit, i.e. when the nonlinear term or potential are large 
in comparison with the kinetic term given by the second-order derivative, 
we can use
two different approximations for describing localized modes. For $\sigma =
+1$ and large {\em
negative} $\Omega$, localised modes are described by the
stationary solutions of the nonlinear Schr\"{o}dinger (NLS) equation, that 
appears when we neglect the trapping potential.  
The NLS one-soliton solution is
$\phi_s (x) = \sqrt{-2\Omega} \, {\rm sech} (x\sqrt{-\Omega})$,
so that the dependence $Q(\Omega)$ coincides with the soliton invariant
$Q_{s} = 4\sqrt{-\Omega}$. For $\sigma =-1$ and large {\em positive}
$\Omega$, the ground-state solution can be obtained by using the so-called
Thomas-Fermi approximation, based on neglecting the kinetic term - this 
yields $\phi_{\rm TF}(x) \approx \sqrt{\Omega-x^2}$.

In general, we should solve Eq. (\ref{GPeq_1}) numerically.  Figures 1(a)
to 1(b) present examples of the numerically  found ground-state solutions
of Eq. (\ref{GPeq_1}) in the form (\ref{local}) as continuous functions of
the dimensionless parameter $\Omega$, for both negative (a) and positive
(b)  scattering length.   For $\Omega \rightarrow 1$, i.e. in the limit of the
harmonic oscillator ground-state mode, the solution is close to Gaussian
for both the cases.  When $\Omega$ deviates from 1, the solution profile is
defined by the  type of nonlinearity.  For attraction ($\sigma = +1$), the
profile
approaches the sech-type soliton, whereas for repulsion  ($\sigma = -1$)
the solution flattens, and it is better described by the Thomas-Fermi
approximation, that is good except at the edge points.

\vspace{-2mm}
\begin{figure}
\setlength{\epsfxsize}{9.0cm}
\centerline{\epsfbox{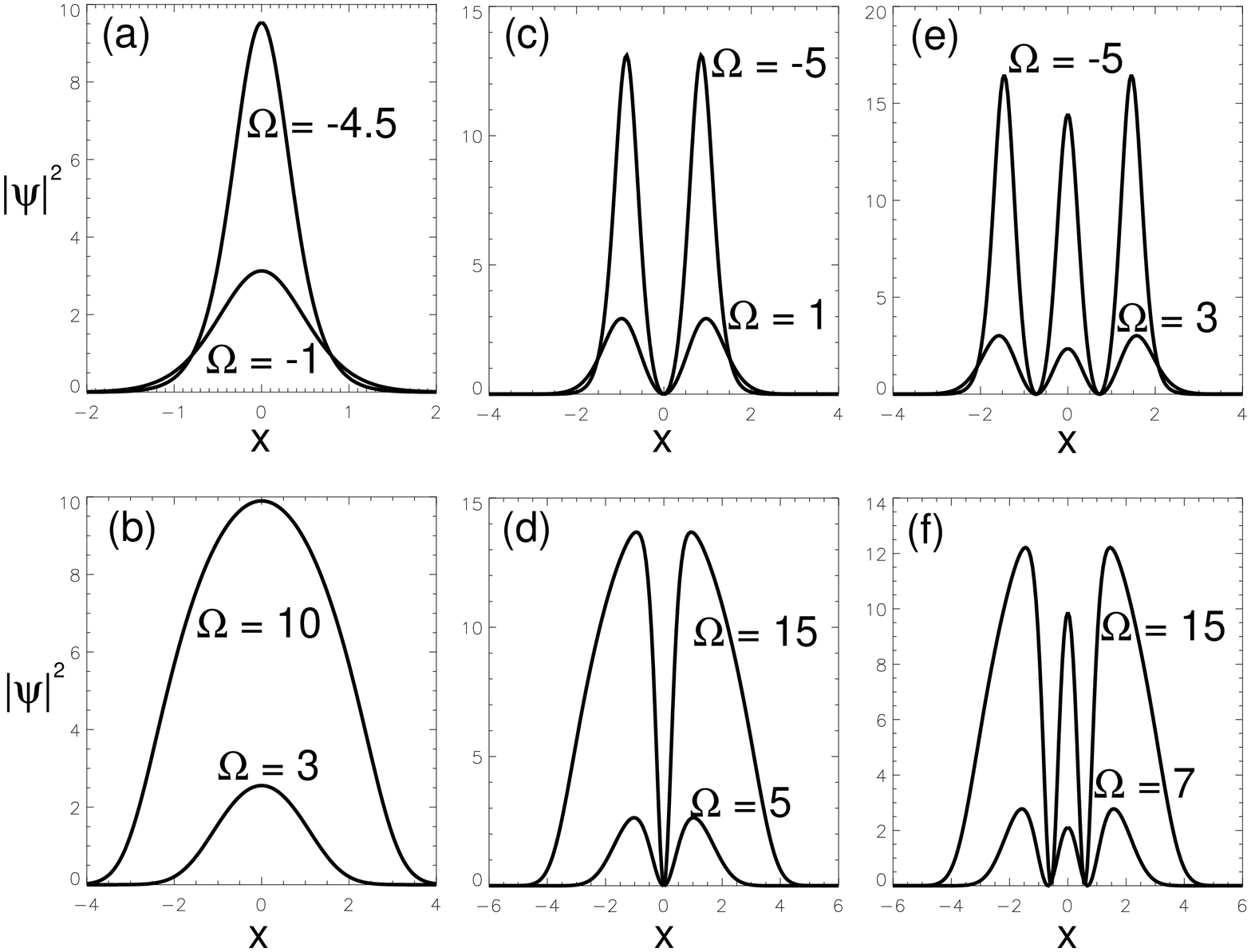}}
\vspace{3mm}
\caption{Examples of the condensate density $|\psi|^{2}$ for the first three
nonlinear modes  described by the stationary solutions of Eq.
(\ref{GPeq_1}) for the negative ($\sigma = +1$, upper row) and positive
($\sigma = -1$, lower row) scattering length, respectively. The values of
$\Omega$ are given next to the curves.}
\label{fig1}
\end{figure}

In Fig. 2 we present the dependence of the invariant $Q$ on the parameter
$\Omega$, for both the types of the ground-state solution, corresponding to
two different signs of the scattering length.  The dashed curve for
the zero-order mode (marked as 0th-mode) shows the  dependence $Q_{\rm
s}(\Omega)$ for the soliton solution  of the NLS equation without a
trapping potential. In the asymptotic region of {\em negative} $\Omega$,
i.e.  say for $\Omega < -2$, the dependence $Q(\Omega)$ for the BEC
condensate in a trap approaches the curve  $Q_{\rm s}(\Omega)$.  This means
that for such a narrow localised state the effect of a parabolic potential
is negligible, and the condensate  ground state becomes localised mostly
due to an attractive interparticle interaction.  In contrast, for large
{\em positive} $\Omega$ the effect of a trapping potential is crucial, and
the solution of the Thomas-Fermi approximation, $\phi_{\rm TF}(x) =
\sqrt{\Omega -x^2}$,  defines the common asymptotics for all the modes, 
$Q_{\rm TF}
\sim \frac{4}{3} \Omega^{3/2}$ (dashed-dotted curves in Fig. 2).

\vspace{-2mm}
\begin{figure}
\setlength{\epsfxsize}{9.0cm}
\centerline{\epsfbox{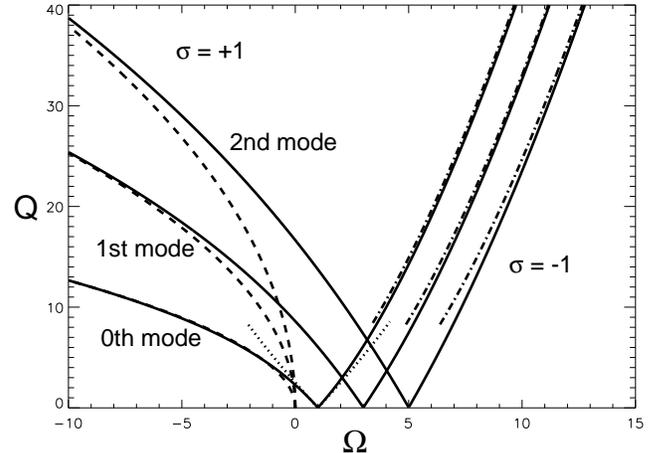}}
\vspace{3mm}
\caption{Invariant $Q$ vs. $\Omega$ for the first three nonlinear modes in
the case of attraction ($\sigma =+1$, left), and repulsion ($\sigma =-1$,
right).
 Dashed curves for $\sigma =+1$ are defined by the soliton invariants
$Q_{s}$, $2Q_{s}$, and $3Q_{s}$, respectively. Dashed-dotted curves for
$\sigma =-1$ are the asymptotic limit given by the Thomas-Fermi
approximation, $Q_{\rm TF} = \frac{4}{3} \Omega^{3/2}$. Dotted lines show the
results of the perturbation theory applied to the ground state, see Eq. 
(\ref{as}).}
\label{fig2}
\end{figure}

As has been mentioned above, in the linear limit ($\sigma \rightarrow 0$),
Eq.  (\ref{GPeq_1}) possesses a discrete set of localised modes described
by the Hermite-Gauss polynomials. We have demonstrated that all such modes
can be readily calculated by the perturbation theory in the weakly nonlinear
approximation, and therefore they should all exist for the nonlinear
problem as well,  describing {\em an analytical continuation of the
Hermite-Gauss linear modes to a set of nonlinear stationary states}.  In 
application to the BEC theory, these non-ground-state solutions were 
first discussed by Yukalov {\em et al} \cite {yukalov}.

Figures 1(c) to 1(f) show examples of the first- and second-order modes for
both negative and positive scattering length, respectively.  In the limit
$\Omega  \rightarrow \Omega_n$, all those modes transform into the
corresponding eigenfunctions of a linear  harmonic oscillator.

It is clear that nonlinearity has a different effect for the  negative
and positive scattering length.  For the negative scattering length
(attraction), the higher-order modes transform into {\em multi-soliton
states}  consisting of a sequence of solitary waves with alternating phases
[see  Figs. 1(c) and 1(e)].  This is further confirmed by the analysis of
the  invariant $Q$ vs. $\Omega$, where all the branches of the higher-order
modes approach asymptotically the soliton dependencies $Q_{n} \sim
(n+1)Q_{\rm s}$, where $n$ is the order of  the mode ($n=0,1,\ldots$). From
the physical point of view, in the case of attractive interaction the
higher-order stationary modes exist due to a balance between  {\em
repulsion} of
out-of-phase bright NLS solitons and {\em attraction} imposed by the trapping
potential. The analysis of  the global stability of such higher-order
multihump multi-soliton modes is still an open problem, however the recent 
results
indicate that, at least in some nonlinear models, multihump soliton states
can be stable \cite{prl}.

For the positive scattering length ($\sigma = -1$), the higher-order modes
transform into a sequence of dark solitons (or {\em kinks}) 
\cite{dark_review}, so that the
first-order  mode corresponds to {\em a single dark soliton}, the
second-order mode, to  {\em a pair of dark solitons}, etc. [see Figs. 1(d)
and 1(f)].  Again, these  stationary solutions satisfy {\em a force balance
condition} - repulsion between  dark solitons is exactly compensated by an
attractive force of the trapping  potential.

The modal structure of the condensate macroscopic states described above
and summarised in Fig. 2 for both positive and negative values of the
scattering length allows us to draw an analogy between BEC in a trap and the
guided-wave optics where the condensate dynamics in time corresponds to the
stationary mode propagation along an optical waveguide, with the parameter
$\Omega$ as the propagation constant.  As is well known from different
problems of guided-wave optics, in the presence of interaction
the guided modes become coupled and the coupling can lead to both power
exchange and nonlinear phase shifting between the modes. In application to
the BEC theory, the mode coupling resembles a kind of internal
Josephson effect. These issues are beyond the scope of this paper and will
be analysed elsewhere \cite{sw}.

The theory of nonlinear stationary modes of a macroscopic quantum
oscillator developed above for the 1D analog of BEC  can be easily extended
to both 2D and 3D cases. Moreover, the coupled-mode theory for a single
condensate is
closely connected to the dynamics of strongly coupled two-component BECs
\cite{doubles} where excitation of an antisymmetric (or, in our notation, the 
first order) collective mode, in the
form of collapses and revivals, has been recently observed experimentally
\cite{two-mode}.

At last, we would like to mention that the basic concepts and results
presented above can find their applications in other fields. For example,
the effect of nonlinearity can lead to a mode coupling in the so-called
``photonic atom'', a micrometer-sized piece of semiconductor that traps
photons inside \cite{ph_atom}, or two such `photonic atoms' coupled
together, a `photonic molecule' \cite{ph_mol}. In such photonic microcavity
structures, the macroscopic nature of the states may lead to different
nonlinear effects including the mode mixing and power exchange.

In conclusion, we have analysed nonlinear stationary modes of a macroscopic
quantum oscillator considering, as an example, the cigar-shaped
Bose-Einstein condensate in a parabolic trap.

\end{multicols}
\end{document}